# Magnetic properties of monoclinic lanthanide metaborates, $Ln(BO_2)_3$, $Ln$ = Pr, Nd, Gd, Tb

P Mukherjee[1*], E Suard[2] and S E Dutton[1*]

[1] Cavendish Laboratory, University of Cambridge, JJ Thomson Avenue, Cambridge CB3 0HE, United Kingdom
[2] Institut Laue-Langevin, 71 Avenue des Martyrs, 38000 Grenoble, France

*E-mail: pm545@cam.ac.uk, sed33@cam.ac.uk

**Abstract**

The bulk magnetic properties of the lanthanide metaborates, $Ln(BO_2)_3$, $Ln$ = Pr, Nd, Gd, Tb are studied using magnetic susceptibility, heat capacity and isothermal magnetisation measurements. They crystallise in a monoclinic structure containing chains of magnetic $Ln^{3+}$ and could therefore exhibit features of low-dimensional magnetism and frustration. $Pr(BO_2)_3$ is found to have a non-magnetic singlet ground state. No magnetic ordering is observed down to 0.4 K for $Nd(BO_2)_3$. $Gd(BO_2)_3$ exhibits a sharp magnetic transition at 1.1 K, corresponding to three-dimensional magnetic ordering. $Tb(BO_2)_3$ shows two magnetic ordering features at 1.05 K and 1.95 K. A magnetisation plateau at a third of the saturation magnetisation is seen at 2 K for both $Nd(BO_2)_3$ and $Tb(BO_2)_3$ which persists in an applied field of 14 T. This is proposed to be a signature of quasi one-dimensional behaviour in $Nd(BO_2)_3$ and $Tb(BO_2)_3$.

# 1 Introduction

Magnetic materials exhibiting geometrical frustration or low-dimensional behaviour have attracted widespread interest due to the unique magnetic properties observed such as the spin ice state in $Ln_2Ti_2O_7$ ($Ln$ = Dy, Ho)[1], dispersionless spin wave excitations in $Gd_3Ga_5O_{12}$[2], order-order transition in $Ca_3Co_2O_6$[3] and quantum critical points (QCPs) in $CoNb_2O_6$[4] and $LiCuSbO_4$[5]. The lattice geometry prevents the magnetic interactions from being satisfied simultaneously in frustrated systems which suppresses, or in the ideal case, completely inhibits magnetic long-range ordering. Ideal one-dimensional magnetic systems also have no long-range order but in bulk materials, inter-chain coupling often leads to complete three-dimensional magnetic ordering. However, such quasi one-dimensional systems still exhibit properties characteristic of one-dimensional magnetism such as magnetisation plateaux and metamagnetism. If the quasi one-dimensional materials are in turn frustrated, exotic magnetic behaviour has been predicted[6–10].

One-dimensional materials containing magnetic lanthanide ($Ln^{3+}$) ions are uncommon; one family of compounds which have recently garnered interest are the lanthanide formates, $Ln(HCOO)_3$. In these compounds, the magnetic $Ln^{3+}$ form one-dimensional chains which are arranged on a triangular lattice. Powder neutron diffraction (PND) experiments on $Tb(HCOO)_3$ indicate development of magnetic order at 1.6 K. This has been ascribed to one-dimensional ferromagnetic order along the chains, analogous to the triangular Ising antiferromagnet state. The direction of chain magnetisation is reported to alternate between neighbouring chains, however the variation is frustrated, inhibiting three-dimensional magnetic ordering[11,12].

Lanthanide metaborates, $Ln(BO_2)_3$, have been well-studied for their applications in phosphors[13–15], however their bulk magnetic properties have not been investigated. They crystallise in two structures depending on the radius of the $Ln^{3+}$ ion – an ambient pressure monoclinic structure for $Ln$ = La – Tb and an orthorhombic structure for $Ln$ = Tb – Er which are synthesised at high pressures [16–19]. Inspection of the reported crystal structure reveals that the magnetic $Ln^{3+}$ form chains in the monoclinic structure. Lanthanide metaborates, $Ln(BO_2)_3$, could therefore be a new example of a $Ln^{3+}$ quasi-one-dimensional magnetic system.

Here we report on the synthesis, characterisation and bulk magnetic properties of monoclinic $Ln(BO_2)_3$ ($Ln$ = Pr, Nd, Gd, Tb). To our knowledge, this is the first report on the bulk magnetic properties of these materials. A broad feature at 5 K is observed in the magnetic susceptibility and heat capacity for $Pr(BO_2)_3$, and no magnetic scattering is observed in low temperature neutron diffraction, indicative of a nonmagnetic singlet ground state. Within the limits of our experiments ($T \geq 0.4$ K), no ordering is observed for $Nd(BO_2)_3$, whereas a sharp magnetic transition is observed for $Gd(BO_2)_3$ at 1.1 K and $Tb(BO_2)_3$ shows two sharp ordering features at 1.05 K and 1.95 K. Both $Nd(BO_2)_3$ and $Tb(BO_2)_3$ show magnetisation plateaus at a third of the saturation magnetisation at 2 K which persist in a field of 14 T. The magnetic properties are discussed in the context of one-dimensional magnetism and frustration.

## 2 Experimental Section

Samples of $Ln(BO_2)_3$ ($Ln$ = Pr, Nd, Gd, Tb) have been prepared by solid state synthesis. Samples were prepared by mixing stoichiometric amounts of $Ln_2O_3$ ($Ln$ = Nd, Gd) (99.999% Alfa Aesar), $Pr_6O_{11}$ (99.999% Alfa Aesar), or $Tb_4O_7$ (99.999% Alfa Aesar) with $H_3BO_3$ (99.99% Alfa Aesar). A 50% excess of $H_3BO_3$ was added to compensate for the loss of B due to volatilisation during heating. $Ln_2O_3$ ($Ln$ = Nd, Gd) and $Pr_6O_{11}$ were pre-dried at 800 ºC prior to being weighed out to ensure the correct stoichiometry. A pre-reaction was carried out at 350 ºC for 2 hours to allow for decomposition of $H_3BO_3$. After regrinding, samples were heated between 800 ºC and 900 ºC for a period of 48 to 192 hours with intermediate regrindings to obtain the final product.

Powder X-Ray diffraction (PXRD) was used to confirm formation of the desired products. Short scans were collected over $10º \leq 2\theta \leq 60º$ ($\Delta 2\theta = 0.015º$) using a Bruker D8 X-Ray diffractometer (Cu K$\alpha$ radiation, $\lambda$ = 1.540 Å). For more quantitative analysis, longer scans for 2 hours over a wide angular range $10º \leq 2\theta \leq 90º$ ($\Delta 2\theta = 0.01º$) were collected. Low temperature PXRD data over the same angular range was collected for $Pr(BO_2)_3$ at 12 K using an Oxford Cryosystems PheniX stage.

For $Ln(BO_2)_3$ ($Ln$ = Pr, Nd, Tb), room temperature (RT) powder neutron diffraction (PND) experiments for structural characterisation were carried out on the D2B diffractometer at ILL, Grenoble ($\lambda$ = 1.595 Å). Samples for the PND experiments were synthesised by the same method as described above but enriched boric acid ($^{11}$B) (99% purity, Sigma Aldrich) was used to reduce the absorption from $^{10}$B. Additional PND measurements were carried out for $Pr(BO_2)_3$ at 3.5 K and 12 K on D2B to investigate any structural transition. Low temperature PND measurements, $T \geq 1.5$ K were carried out on $Pr(BO_2)_3$ on the D1B diffractometer at ILL, Grenoble ($\lambda$ = 2.525 Å) to investigate the existence of any magnetic ordering. PXRD and PND Rietveld refinements were carried out using the Fullprof suite of programs[20]. The background was modelled using linear interpolation, and peak shape using a pseudo-Voigt function. The appropriate scattering length for enriched boron was used in the PND Rietveld refinement.

Magnetic measurements were made on a Quantum Design Magnetic Properties Measurement System (MPMS) with a Superconducting Quantum Interference Device (SQUID) magnetometer. The zero field cooled (ZFC) susceptibility $\chi(T)$ was measured in a field of 100 Oe in the temperature range 2 – 300 K for all $Ln(BO_2)_3$ ($Ln$ = Pr, Nd, Gd, Tb). In a low field of 100 Oe, the isothermal magnetisation $M(H)$ curve is linear at all $T$ and so the linear approximation for $\chi(T)$ is valid, that is, $\chi(T) \sim M/H$. $M(H)$ measurements in the field range, $\mu_o H = 0 – 9$ T were carried out on all samples at selected temperatures using the ACMS (AC Measurement System) option on a Quantum Design Physical Properties Measurement System (PPMS).

Zero field heat capacity (HC) measurements for $Ln(BO_2)_3$ ($Ln$ = Pr, Nd, Gd, Tb) were performed using a Quantum Design PPMS in the temperature range 0.4 – 20 K using the He3 option. To improve thermal conductivity at low temperatures, samples were mixed with approximately equal mass of silver powder (99.99%, Alfa Aesar). The contribution of the silver to the heat capacity was then deducted[21] to obtain the heat capacity of the sample.

The lattice contribution was subtracted using a Debye model[22] with $\theta_D$ = 325 K for Pr(BO$_2$)$_3$, $\theta_D$ = 315 K for Nd(BO$_2$)$_3$, $\theta_D$ = 275 K for Gd(BO$_2$)$_3$ and $\theta_D$ = 250 K for Tb(BO$_2$)$_3$ to obtain the magnetic contribution to the heat capacity, $C_m(T)$.

# 3 Results

## 3.1 Crystal Structure

PXRD indicated formation of $Ln$(BO$_2$)$_3$, $Ln$ = Pr, Nd, Gd, Tb, with minor impurities of $Ln$BO$_3$ (< 2% by weight) and unreacted H$_3$BO$_3$ (< 2% by weight) detected (Table 1). Attempts to synthesise $Ln$(BO$_2$)$_3$ ($Ln$ = Dy, Ho, Er) were unsuccessful. A mixture of $Ln$(BO$_2$)$_3$ and $Ln$BO$_3$ was obtained for Dy while only the $Ln$BO$_3$ phase was obtained for Ho and Er. We conclude that the monoclinic phase cannot be obtained for $Ln$(BO$_2$)$_3$ ($Ln$ = Dy, Ho, Er) by this synthesis method. This is consistent with earlier reports where a high pressure synthesis was required for $Ln$(BO$_2$)$_3$, $Ln$ = Dy – Er, and only the orthorhombic phase could be obtained[19].

Lanthanide metaborates, $Ln$(BO$_2$)$_3$, $Ln$ = Pr, Nd, Gd, Tb, crystallise in a monoclinic unit cell with space group $C2/c$. Ribbons of borate units comprising alternating corner sharing BO$_3^{3-}$ triangles and BO$_4^{5-}$ tetrahedra propagate along the $c$ axis (Figure 1a). The $Ln^{3+}$ ions occupy sites between the ribbons forming one-dimensional chains (Figure 1b). The interactions between chains are complex, with the $Ln^{3+}$ ions forming a distorted three-dimensional honeycomb lattice (Figure 1c).

The crystal structure for $Ln$(BO$_2$)$_3$ was refined from previous reports [17,23,24] by Rietveld refinement of the RT PND and PXRD patterns for $Ln$ = Pr, Nd, Tb and from the PXRD alone for $Ln$ = Gd. PND is sensitive to the B and O positions and so a combined refinement gives accurate information about the crystal structure. The boron content in the structure was refined and found to be stoichiometric within error, the B occupancy was therefore fixed to be consistent with the $Ln$(BO$_2$)$_3$ stoichiometry. The combined PXRD + PND refinements for Pr(BO$_2$)$_3$ are shown in Figure 2 and the refined parameters are compiled in Table 1. The lattice parameters $a$ and $c$ are found to change linearly with the $Ln^{3+}$ ionic radius[25] while $b$ and $\beta$ deviate slightly from the linear trend (Figure 3). However they compensate each other and overall the lattice volume varies linearly with the $Ln^{3+}$ ionic radius.

The low-temperature crystal structure for Pr(BO$_2$)$_3$ was refined from a combined PXRD and PND refinement at 12 K. There is a decrease in the lattice parameters and slight shifts in the atomic positions on cooling (Table 2); however the monoclinic crystal structure is retained. No further structural changes are detected between 3.5 K and 12 K from the PND data. We conclude that there is no structural transition at low temperatures in Pr(BO$_2$)$_3$.

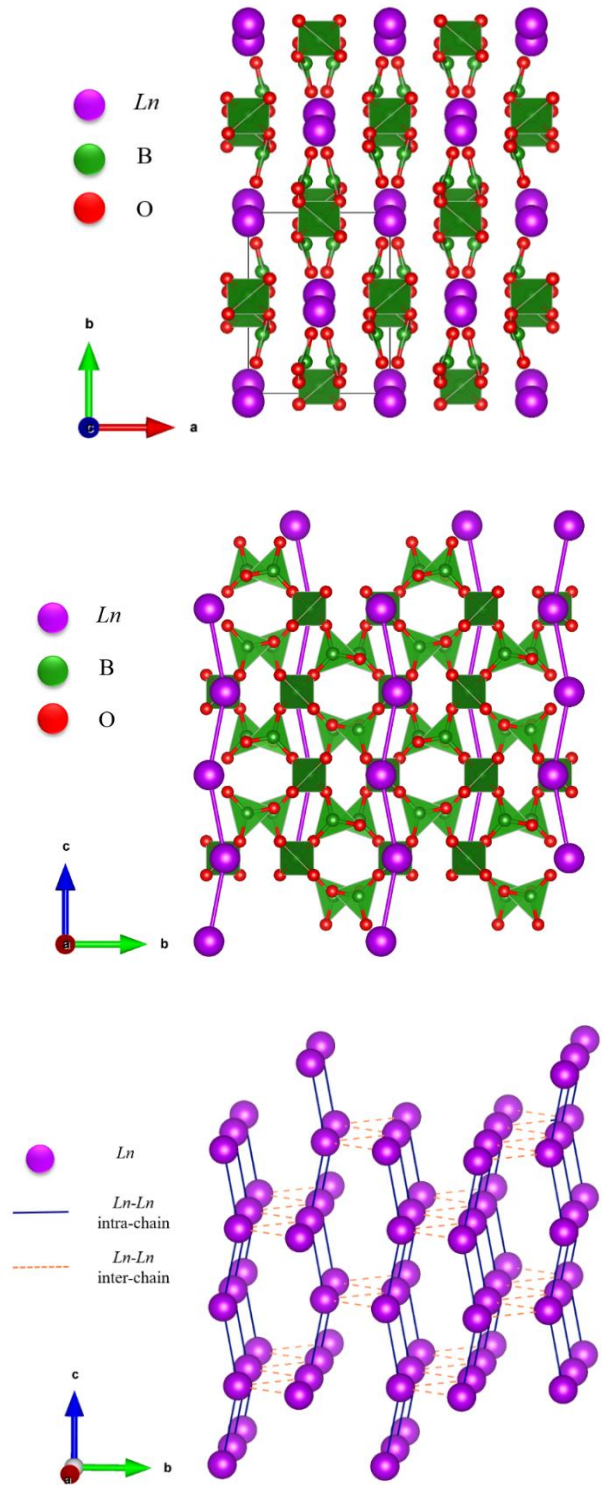

Figure 1 – General crystal structure for monoclinic $Ln(BO_2)_3$ – a) Ribbons of borate groups consisting of $(BO_4)^{5-}$ tetrahedra and $(BO_3)^{3-}$ triangles propagate along the *c* axis b) Magnetic $Ln^{3+}$ form one-dimensional chains between the ribbons c) If inter-chain dipolar interactions are considered along with intra-chain interactions, the magnetic $Ln^{3+}$ form a distorted honeycomb lattice in three dimensions

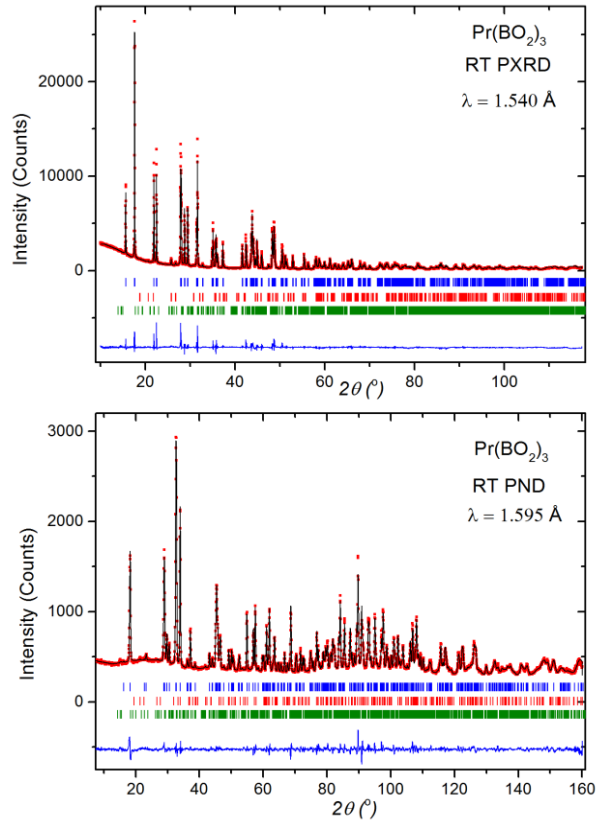

Figure 2 – Room temperature PXRD and PND patterns for Pr(BO$_2$)$_3$: Experimental data (red dots), Modelled data (black line), Difference pattern (blue line), Bragg positions: (blue ticks) - Pr(BO$_2$)$_3$, (red ticks) – PrBO$_3$, (green ticks) – H$_3$BO$_3$.

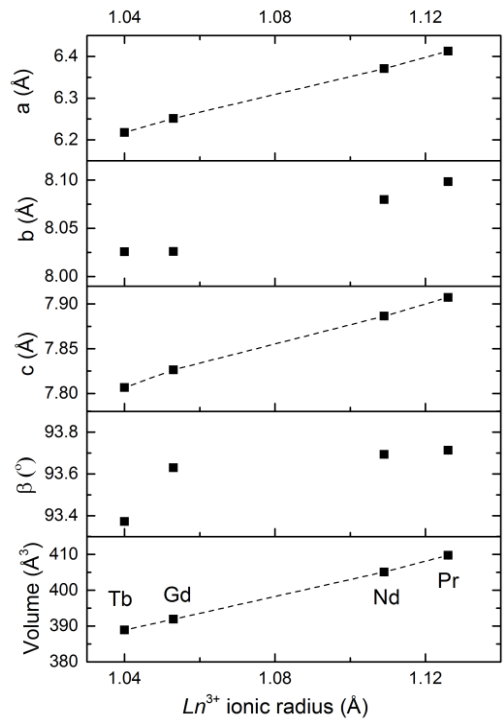

Figure 3 – Plot of lattice parameters for $Ln$(BO$_2$)$_3$ as a function of $Ln^{3+}$ ionic radii: Lines are a guide to the eye

Table 1 – Crystal structure parameters for monoclinic $Ln(BO_2)_3$ - space group $C2/c$.

| $Ln$ | | Pr | Nd | Gd [a] | Tb |
|---|---|---|---|---|---|
| $a$ (Å) | | 6.41249 (3) | 6.37043 (6) | 6.25159 (5) | 6.21781 (5) |
| $b$ (Å) | | 8.09827 (6) | 8.07975 (9) | 8.02602 (9) | 8.02564 (6) |
| $c$ (Å) | | 7.90730 (5) | 7.88632 (9) | 7.82649 (8) | 7.80659 (4) |
| $\beta$ (Å) | | 93.7129 (5) | 93.6935 (7) | 93.6292 (8) | 93.3723 (4) |
| Volume (Å$^3$) | | 409.765 (5) | 405.078 (7) | 391.909 (7) | 388.889 (6) |
| $\chi^2$ | | 5.71 | 5.00 | 4.05 | 3.51 |
| $R_{wp}$ | | 3.58 | 3.32 | 5.72 | 2.66 |
| $Ln$: $4e$ $(0, y, ¼)$ | $y$ | 0.0501 (2) | 0.0499 (2) | 0.0464 (2) | 0.04500 (13) |
| | $B_{iso}$ (Å$^2$) | 0.61 (5) | 1.55 (3) | 0.5 [b] | 0.76 (5) |
| B1: $4e$ $(0, y, ¼)$ | $y$ | 0.4714 (4) | 0.4699 (5) | 0.4590 [b] | 0.4728 (4) |
| | $B_{iso}$ (Å$^2$) | 0.41 (4) | 0.53 (8) | 0.5 [b] | 1.37 (8) |
| B2: $8f$ $(x, y, z)$ | $x$ | 0.1074 (2) | 0.1066 (5) | 0.09800 [b] | 0.1095 (2) |
| | $y$ | 0.3166 (2) | 0.3164 (3) | 0.32600 [b] | 0.3137 (2) |
| | $z$ | 0.5249 (2) | 0.5256 (4) | 0.52100 [b] | 0.5248 (2) |
| | $B_{iso}$ (Å$^2$) | 0.18 (2) | 0.23 (5) | 0.5 [b] | 0.14 (2) |
| O1: $8f$ $(x, y, z)$ | $x$ | 0.1414 (2) | 0.1430 (5) | 0.15045 [b] | 0.1493 (3) |
| | $y$ | 0.3629 (2) | 0.3605 (4) | 0.36671 [b] | 0.3621 (2) |
| | $z$ | 0.3578 (2) | 0.3571 (4) | 0.35116 [b] | 0.3553 (2) |
| | $B_{iso}$ (Å$^2$) | 0.52 (3) | 0.58 (5) | 0.5 [b] | 0.97 (3) |
| O2: $8f$ $(x, y, z)$ | $x$ | 0.1395 (3) | 0.1407 (5) | 0.14316 [b] | 0.1455 (2) |
| | $y$ | 0.4368 (2) | 0.4362 (4) | 0.43845 [b] | 0.4351 (3) |
| | $z$ | 0.6479 (3) | 0.6480 (4) | 0.64816 [b] | 0.6508 (3) |
| | $B_{iso}$ (Å$^2$) | 0.70 (3) | 0.47 (5) | 0.5 [b] | 0.75 (5) |
| O3: $8f$ $(x, y, z)$ | $x$ | 0.0479 (3) | 0.0491 (5) | 0.05526 [b] | 0.0522 (4) |
| | $y$ | 0.1617 (3) | 0.1601 (5) | 0.17006 [b] | 0.1582 (4) |
| | $z$ | 0.5472 (3) | 0.5476 (5) | 0.55122 [b] | 0.5434 (4) |
| | $B_{iso}$ (Å$^2$) | 0.18 (2) | 0.93 (7) | 0.5 [b] | 1.30 (7) |
| $Ln$BO$_3$ wt % | | 1.3 (4) | 1.8 (4) | 0.4 (5) | 1.8 (3) |
| H$_3$BO$_3$ wt % | | 0.56 (14) | 1.82 (17) | - [c] | 0.90 (14) |
| $Ln$ –O1 (Å) | | 2.602 (3) × 2 | 2.580 (4) × 2 | 2.472 (9) × 2 | 2.489 (4) × 2 |
| | | 2.801 (3) × 2 | 2.781 (4) × 2 | 2.832 (9) × 2 | 2.815 (4) × 2 |
| $Ln$ –O2 (Å) | | 2.568 (3) × 2 | 2.548 (4) × 2 | 2.477 (9) × 2 | 2.457 (6) × 2 |
| $Ln$ –O3 (Å) | | 2.383 (4) × 2 | 2.364 (5) × 2 | 2.373 (10) × 2 | 2.330 (4) × 2 |
| | | 2.520 (4) × 2 | 2.510 (4) × 2 | 2.561 (10) × 2 | 2.467 (4) × 2 |
| <$Ln$ – O> (Å) | | 2.575 | 2.557 | 2.543 | 2.512 |

[a] Parameters determined from PXRD only.
[b] Not refined, positions taken from previous structural report[24].
[c] Could not be obtained from PXRD.

Table 2

Comparison of structural information for Pr(BO$_2$)$_3$ from combined PXRD + PND refinements at 12 K and 300 K (RT).

| | | 12 K | 300 K |
|---|---|---|---|
| *a* (Å) | | 6.40092 (6) | 6.41249 (3) |
| *b* (Å) | | 8.08400 (4) | 8.09827 (6) |
| *c* (Å) | | 7.90419 (7) | 7.90730 (5) |
| *β* ( Å) | | 93.9367 (7) | 93.7129 (5) |
| **Volume (Å$^3$)** | | 408.038 (6) | 409.765 (5) |
| $\chi^2$ | | 5.90 | 5.71 |
| **R$_{wp}$** | | 3.61 | 3.58 |
| *Ln*: *4e* (0, *y*, ¼) | *y* | 0.0509 (5) | 0.05007 (20) |
| | B$_{iso}$ (Å$^2$) | 0.38 (6) | 0.61 (5) |
| **B1**: *4e* (0, *y*, ¼) | *y* | 0.4712 (3) | 0.4714 (4) |
| | B$_{iso}$ (Å$^2$) | 0.21 (5) | 0.41 (4) |
| **B2**: *8f* (*x*, *y*, *z*) | *x* | 0.1072 (3) | 0.10745 (24) |
| | *y* | 0.31634 (22) | 0.31665 (18) |
| | *z* | 0.52479 (24) | 0.52493 (20) |
| | B$_{iso}$ (Å$^2$) | 0.16 (3) | 0.183 (23) |
| **O1**: *8f* (*x*, *y*, *z*) | *x* | 0.1423 (3) | 0.14141 (25) |
| | *y* | 0.36283 (27) | 0.36293 (22) |
| | *z* | 0.35596 (28) | 0.35785 (24) |
| | B$_{iso}$ (Å$^2$) | 0.27 (4) | 0.52 (3) |
| **O2**: *8f* (*x*, *y*, *z*) | *x* | 0.1397 (3) | 0.1395 (3) |
| | *y* | 0.43637 (26) | 0.43675 (24) |
| | *z* | 0.6481 (3) | 0.64796 (27) |
| | B$_{iso}$ (Å$^2$) | 0.36 (4) | 0.70 (3) |
| **O3**: *8f* (*x*, *y*, *z*) | *x* | 0.0467 (3) | 0.04794 (28) |
| | *y* | 0.1609 (3) | 0.16173 (28) |
| | *z* | 0.5477 (3) | 0.54722 (28) |
| | B$_{iso}$ (Å$^2$) | 0.41 (4) | 0.183 (23) |

## 3.2 Bulk Magnetic Properties

Magnetic susceptibility, $\chi(T)$, zero field magnetic heat capacity, $C_m(T)$ and isothermal magnetisation $M(H)$ measurements are shown in Figures 4, 5 and 6 respectively. For all samples, the reciprocal susceptibility $\chi^{-1}$ is linear above 100 K and was used to fit to the Curie-Weiss law, $\chi = \frac{C}{T-\theta_{CW}}$ where $C$ is the Curie constant and $\theta_{CW}$ is the Curie-Weiss temperature. Parameters for the Curie-Weiss fits are summarised in Table 3. The Curie-Weiss temperatures are negative for all samples indicating antiferromagnetic interactions. Significant temperature-independent paramagnetism (TIP) is observed in Nd(BO$_2$)$_3$ and so the high $T$ fit to the Curie-Weiss law gives an unrealistically large value for $\theta_{CW}$, and hence is not reported. Instead, as has been reported for other frustrated Nd$^{3+}$ systems[26–30], we consider the low T regime, 2 – 30 K. A fit in this temperature range gives $\mu_{eff}$ = 2.14 $\mu_B$ and $\theta_{CW}$ = -0.20 K. The experimentally determined magnetic moments generally agree well with the theoretical values.

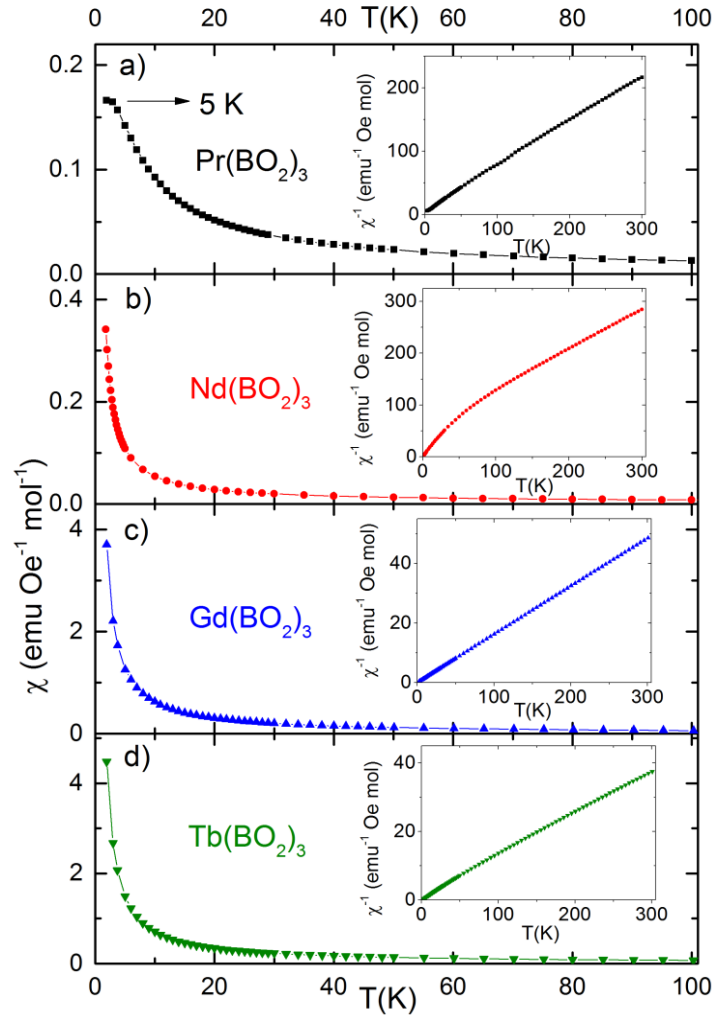

Figure 4 - Zero field cooled (ZFC) magnetic susceptibility $\chi(T)$ measured in a field of 100 Oe for $Ln$(BO$_2$)$_3$; the inverse susceptibility $\chi^{-1}(T)$ is inset

Table 3 - Magnetic properties of monoclinic $Ln(BO_2)_3$

| $Ln$ | $T_N$ (K) | $\theta_{CW}$ (K) | Theoretical | | Experimental | |
|---|---|---|---|---|---|---|
| | | | $\mu_{eff}$ ($\mu_B$) | $M_{sat} = g_J J$ ($\mu_B$/f.u.) | $\mu_{eff}$ ($\mu_B$) | $M_{max2K,9T}$ ($\mu_B$/f.u.) |
| **Pr** | - | -14 (3) | 3.58 | 3.2 | 3.38 (3) | 1.3 |
| **Nd** | <0.4 | TIP [a] | 3.62 | 3.3 | 3.15 (7) | 1.0 |
| **Gd** | 1.1 | -0.4 (2) | 7.94 | 7.0 | 7.032 (5) | 6.0 |
| **Tb** | 1.05, 1.95 | -13 (3) | 9.72 | 9.0 | 8.11 (7) | 2.9 |

[a] Low temperature fit from 2 – 30 K gives $\mu_{eff}$ = 2.14 $\mu_B$ and $\theta_{CW}$ = -0.20 K. This value of $\theta_{CW}$ has been used for calculating $J_1$ in Table 3.

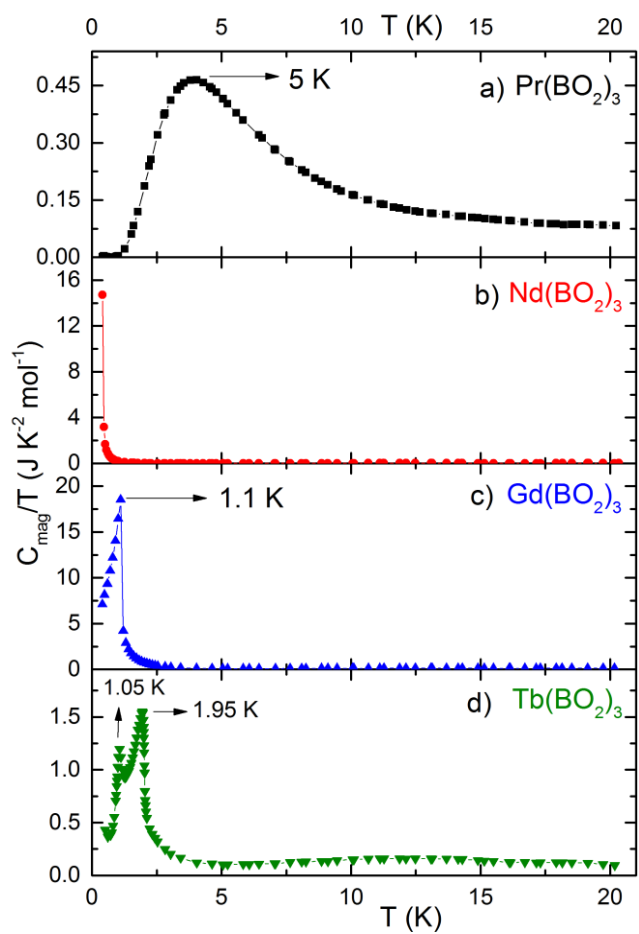

Figure 5 – Zero field magnetic heat capacity $C_m(T)$ for $Ln(BO_2)_3$

With the exception of $Pr(BO_2)_3$ no features in the magnetic susceptibility or heat capacity are observed above 2 K. $Pr(BO_2)_3$ shows a broad feature at 5 K in the magnetic susceptibility (Figure 4a) and heat capacity (Figure 5a). No additional ordering transitions are observed below 2 K in the specific heat. The isothermal magnetisation shows no signs of saturation in a field of 9 T (Figure 6a). Low temperature PND of $Pr(BO_2)_3$ shows no magnetic Bragg peaks or diffuse scattering down to 1.5 K (Figure 7). We conclude that it has a non-magnetic singlet ground state as has been reported for other $Pr^{3+}$ containing samples[30,31], the low symmetry crystal structure are consistent with this hypothesis. The broad feature at 5 K in the bulk magnetic measurements is attributed to van Vleck paramagnetism, as has been reported elsewhere[31].

Zero field heat capacity measurements below 2 K indicate magnetic transitions for all the other lanthanide metaborates studied, with the exception of Nd(BO$_2$)$_3$ where the onset of a sharp transition can be seen beyond the temperature limit of our measurements (0.4 K). Two sharp magnetic transitions at 1.05 K and 1.95 K are seen in Tb(BO$_2$)$_3$ whereas a single sharp λ type transition at 1.1 K is observed for Gd(BO$_2$)$_3$.

Isothermal magnetisation measurements for all three paramagnetic metaborates saturate at high fields at lower temperatures, $T \leq 4$ K. Gd(BO$_2$)$_3$ saturates at 6 $\mu_B$/f.u in a field of 9 T, close to the saturation value, $M_{sat}$, expected for Heisenberg spins ($M_{sat} = g_J J = 7$ $\mu_B$/f.u for Gd$^{3+}$). Both Nd(BO$_2$)$_3$ and Tb(BO$_2$)$_3$ saturate at values close to ~ $M_{sat}/3$. Measurements to higher fields show the persistence of the plateau at $M_{sat}/3$ in the limiting field of $\mu_o H = 14$ T (Figure S1).

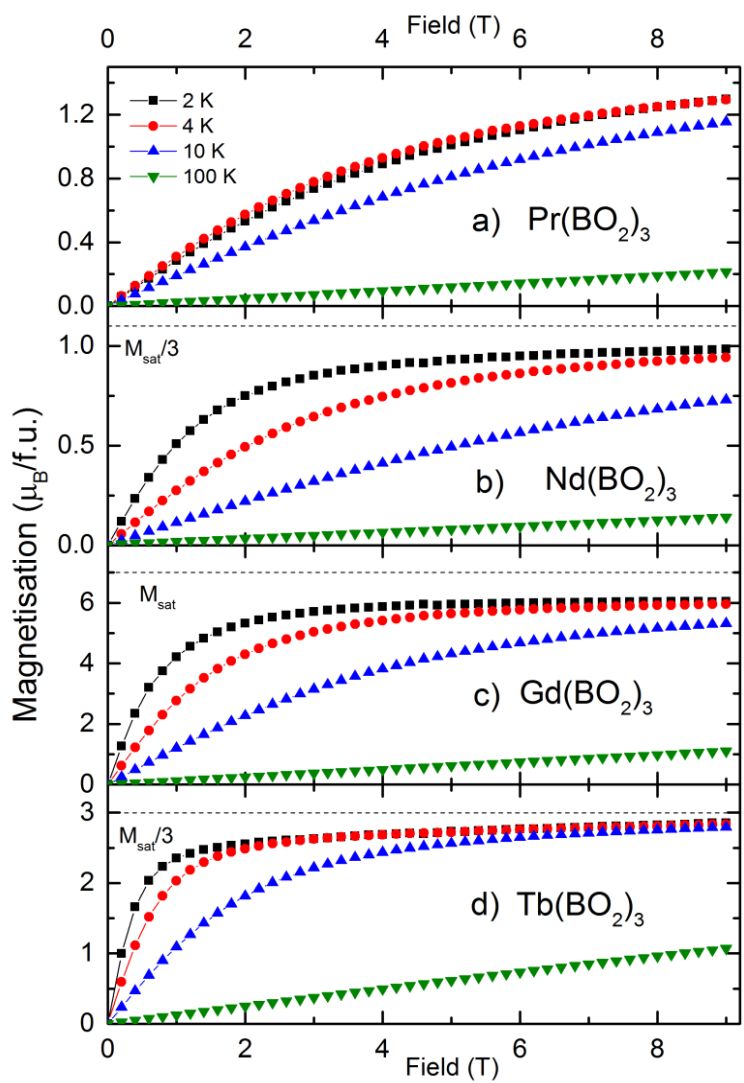

Figure 6 – Isothermal magnetisation $M(H)$ curves at selected temperatures for $Ln$(BO$_2$)$_3$

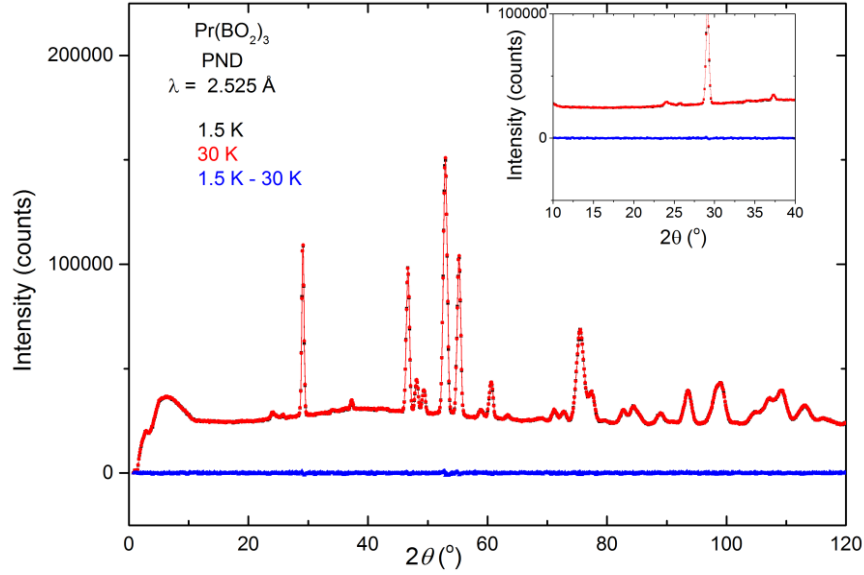

Figure 7 – PND pattern for Pr(BO$_2$)$_3$ on D1B, ILL at 1.5 K and 30 K and the difference plot; inset: a zoom-in of the same plot at low 2$\theta$; note the absence of any magnetic Bragg or diffuse scattering

## 3.3 Discussion

We now consider the bulk magnetic properties of the monoclinic lanthanide metaborates in the context of one-dimensional magnetism and frustration. Pr(BO$_2$)$_3$ has a non-magnetic singlet ground state, this is likely due to crystal electric field (CEF) effects that lead to the ground state being a well-isolated singlet as reported for other Pr$^{3+}$ systems [30,31]. We thus exclude Pr(BO$_2$)$_3$ from this discussion and focus on $Ln$ = Nd, Gd, Tb.

Table 4 – Dipolar ($D$) and nearest-neighbour exchange ($J_1$) interactions for monoclinic $Ln$(BO$_2$)$_3$, $Ln$ = Nd, Gd, Tb; $Ln$-$Ln$ distances are obtained from structural PXRD and PND analysis.

| $Ln$ | Nd | Gd | Tb |
| --- | --- | --- | --- |
| $Ln$ – $Ln$ intra-chain (Å) | 4.0241 (5) | 3.9838 (5) | 3.9696 (4) |
| $D_{\text{intra-chain}}$ (K) | 0.005 | 0.04 | 0.022 |
| $Ln$ – $Ln$ inter-chain (Å) | 4.5374 (17) | 4.5212 (17) | 4.5269 (15) |
| $D_{\text{inter-chain}}$ (K) | 0.003 | 0.03 | 0.015 |
| $J_1$ (K) | $S = 3/2$ : -0.04 | $S = 7/2$: -0.02 | $S = 3$ : -0.8 |
|  | $S = 1/2$ : -0.2 |  | $S = 1/2$ : -12.5 |

The magnetic behaviour is determined by the connectivity of the magnetic $Ln^{3+}$ and the competition among the different magnetic interactions. In addition to the superexchange, the dipolar interaction also needs to be considered at low temperatures as the trivalent rare-earth ions have large magnetic moments[32–34]. CEF effects will also have a significant impact on the single-ion anisotropy and the magnetic properties for $Ln^{3+}$ with non-zero values of orbital angular momentum[35,36]. For Nd(BO$_2$)$_3$ and Tb(BO$_2$)$_3$, such effects could lead to the ground state doublet being well isolated at low temperatures with $S = ½$ and Ising behaviour as reported for other systems [29,37,38]. Accurate modelling of all the relevant interactions require further inelastic neutron scattering experiments, however an order of magnitude approximation can be obtained from the bulk magnetic data as follows: The scale for the dipolar interaction can be estimated as $D \cong \frac{\mu_0 g_L^2 \mu_B^2}{4\pi R_{nn}^3}$, where $R_{nn}$ is the distance between neighbouring $Ln^{3+}$ and $g_L$ is the Landé factor[33–35,39]. The Curie-Weiss constant $\theta_{CW}$ contains contributions from the nearest-neighbour exchange as well as other terms[32,40,41]. However, one can obtain an approximate value for the nearest-neighbour isotropic exchange using the mean-field result[40–42] as $J_1 \cong \frac{3k_B \theta_{CW}}{2nS(S+1)}$ where $n$ is the number of nearest-neighbour $Ln^{3+} = 2$ and $S$ is the spin quantum number.

We propose that the relative magnitude of the $J_1$ and $D$ interactions determine whether these materials exhibit quasi one-dimensional behaviour. The inter ($D_{inter-chain}$) and intra-chain ($D_{intra-chain}$) dipolar interactions are of comparable magnitude ($D_{inter-chain}/D_{intra-chain} \sim 0.7$) and so, considering predominantly dipolar interactions, the magnetic $Ln^{3+}$ form a distorted hyper honeycomb lattice (Figure 1c) and are expected to show three-dimensional magnetic ordering. If the $J_1$ interactions dominate, quasi one-dimensional magnetic behaviour may still be observed.

Frustration in quasi one-dimensional systems can arise due to competition between nearest neighbour ($J_1$) and next nearest neighbour exchange interactions ($J_2$). In general, it is highly dependent on the single-ion anisotropy of the magnetic $Ln^{3+}$ ion and the relative magnitude of competing interactions; this has been observed in the case of other series of lanthanide containing geometrically frustrated magnets like the pyrochlores [32]. However a quantitative estimate can be obtained by considering the criterion proposed by Ramirez which states that a material is strongly frustrated if the frustration index, $f = \left|\frac{\theta_{CW}}{T_0}\right|$, is > 10, where $T_0$ is the ordering temperature[42]. Using this, we obtain $f = 0.4$ for Gd(BO$_2$)$_3$ and $f = 11.9$ and $6.4$ corresponding to the two transitions for Tb(BO$_2$)$_3$.

For Gd(BO$_2$)$_3$ the dipolar interactions are marginally greater than the $J_1$ exchange interactions. The ordering transition ($T_N = 1.1$ K) is very close to the Curie-Weiss temperature, as expected for an antiferromagnet with no geometric frustration. The isothermal magnetisation is consistent with the Heisenberg nature of the Gd$^{3+}$ spins. Thus we postulate that the sharp λ type transition in the zero field heat capacity at 1.1 K corresponds to three-dimensional ordering of a Heisenberg antiferromagnet.

For $Ln$ = Nd and Tb, the $J_1$ interactions, calculated for both the usual spin quantum number, and considering them as effective $S = 1/2$ systems, are much greater than both $D_{intra-chain}$ and $D_{inter-chain}$; hence the magnetic behaviour would be quasi one-dimensional in nature. This is consistent with the bulk magnetic measurements; specifically the observation of a plateau at 1/3 of the saturation magnetisation at 2 K, for both Nd(BO$_2$)$_3$ and Tb(BO$_2$)$_3$. We propose that this corresponds to a field-induced transition similar to that in other quasi one-dimensional Ising compounds like CoV$_2$O$_6$ and CoNb$_2$O$_6$ [43,44]. However unlike those systems, the plateau persists in a field of 14 T. Also unusually, the magnetisation plateaux are observed above the magnetic ordering for both $Ln$ = Nd and Tb. We postulate that this is due to existence of short-range magnetic correlations above the long-range ordering transitions. Tb(BO$_2$)$_3$ has two sharp transitions at 1.05 K and 1.95 K in zero field. This is reminiscent of the behaviour observed in quasi one-dimensional CoNb$_2$O$_6$ which has two transitions at 1.97 K and 2.97 K, corresponding to a superposition of two commensurate phases and an incommensurate magnetic structure respectively[45]. The ordering transitions lie well below the Curie-Weiss temperature, implying possible frustration in Tb(BO$_2$)$_3$.

For Nd(BO$_2$)$_3$ no magnetic ordering is seen down to 0.4 K although the sharp increase in $C_{mag}/T$ at 0.4 K can be attributed to the onset of an ordering transition. The weaker $D$ and $J_1$ interactions are postulated to lead to magnetic ordering at lower temperatures as compared to Tb(BO$_2$)$_3$. The low temperature Curie-Weiss fit at lower temperatures for Nd(BO$_2$)$_3$ gives $\theta_{CW}$ = -0.2 K and no conclusions can be drawn regarding the frustration as the ordering transition is beyond the temperature limit of our measurements.

Further work such as zero field PND measurements at $T < T_0$, inelastic neutron scattering experiments and PND measurements in field are required to determine the magnetic ground state and precisely model the magnetic interactions, single-ion anisotropy and field-induced transitions in $Ln$(BO$_2$)$_3$. This would provide a more complete understanding of the origin of the quasi one-dimensional magnetic behaviour and frustration in these materials. Further, the lanthanide metaborates could also find practical applications as low temperature magnetocaloric materials, $T \geq 2$ K, as has been reported for the lanthanide formates[11].

# 4 Conclusion

We have synthesised polycrystalline samples of $Ln$(BO$_2$)$_3$, $Ln$ = Pr, Nd, Gd, Tb, characterised the structure using RT PXRD and PND and studied their bulk magnetic properties. The RT Rietveld refinements show that they crystallise in a monoclinic structure, consistent with previous reports. It is found that Pr(BO$_2$)$_3$ has a non-magnetic singlet ground state. Gd(BO$_2$)$_3$ undergoes antiferomagnetic ordering at 1.1 K and the magnetisation saturates at 6 $\mu_B$/f.u, consistent with isotropic Gd$^{3+}$ spins. The onset of magnetic ordering is seen in Nd(BO$_2$)$_3$ at 0.4 K while Tb(BO$_2$)$_3$ shows two sharp magnetic ordering features at 1.05 K and 1.95 K. Both Nd(BO$_2$)$_3$ and Tb(BO$_2$)$_3$ show a $M_{sat}/3$ magnetisation plateau in fields up to 14 T at $T =$ 2 K, consistent with quasi one-dimensional behaviour seen in other Ising compounds like CoV$_2$O$_6$ and CoNb$_2$O$_6$.

The monoclinic $Ln$(BO$_2$)$_3$, $Ln$ = Pr, Nd, Gd, Tb exhibit a wide range of interesting magnetic behaviour. We hope that this work will motivate further investigations into the magnetic properties of these hitherto unexplored materials.

# 5 Acknowledgements

We thank J. Hodkinson for his support during the PND experiments on D1B and D2B and for valuable feedback. We acknowledge funding support from the Winton Programme for the Physics of Sustainability. Magnetic measurements were carried out using the Advanced Materials Characterisation Suite, funded by EPSRC Strategic Equipment Grant EP1M00052411. Supporting data can be found at https://doi.org/10.17863/CAM.11956, neutron diffraction data can also be found at doi:10.5291/ILL-DATA.5-31-2456.

# 7 Supplementary Information

Figure S1

Magnetisation in a field of 14 T for Nd(BO$_2$)$_3$ and Tb(BO$_2$)$_3$; in both cases the $M_{sat}/3$ plateau is found to persist.

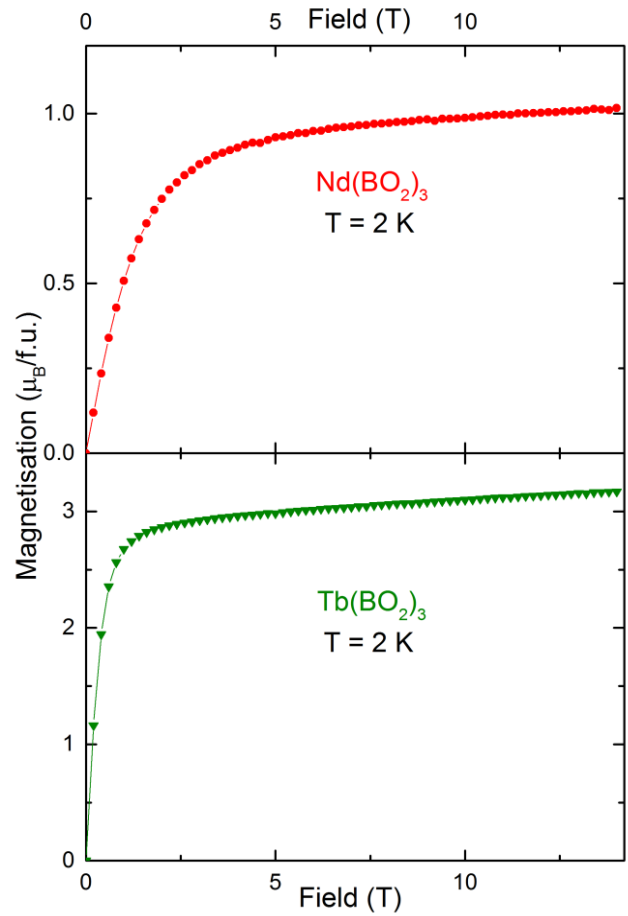

Figure S2

Fits to the inverse susceptibility in the temperature range 100 – 300 K for $Ln(BO_2)_3$, $Ln$ = Pr, Gd, Tb and in the temperature range 2 – 30 K for $Nd(BO_2)_3$

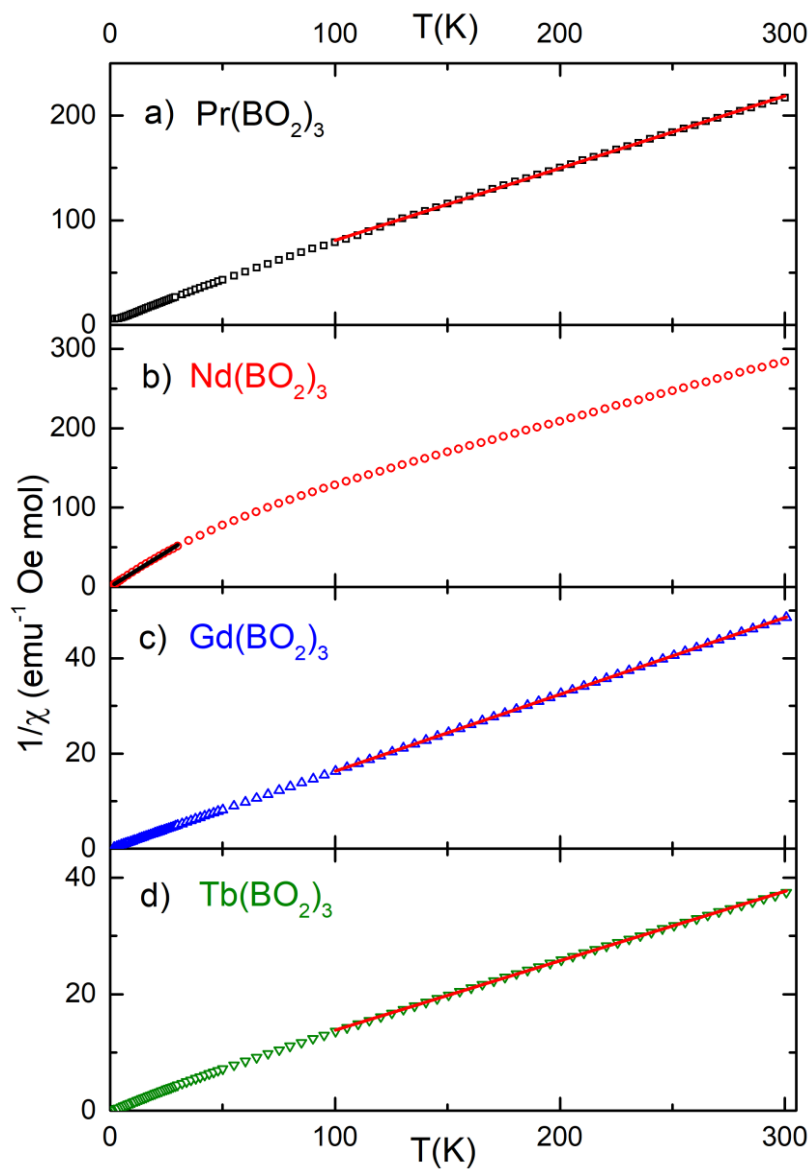